\documentclass[onecolumn]{aastex631}
\usepackage{longtable}

\usepackage{graphicx}
\usepackage{xcolor}
\usepackage{natbib}
\usepackage{amssymb,amsmath}
\usepackage{CJKutf8}

\def \NAU {Department of Astronomy and Planetary Science, Northern Arizona University,\\PO Box 6010, Flagstaff, AZ 86011, USA}
\def \UMPhysics {Department of Physics, University of Michigan,\\ Ann Arbor, MI 48109, USA}
\def \UMAstronomy {Department of Astronomy, University of Michigan,\\ Ann Arbor, MI 48109, USA}
\def \UW {DiRAC Institute and the Department of Astronomy, University of Washington, Seattle, USA}
\def \uchile {Departamento de Astronomía, Universidad de Chile,\\ Camino del Observatorio 1515, Las Condes, Santiago, Chile}
\def \cfa {Harvard-Smithsonian Center for Astrophysics,\\ 60 Garden St., MS 51, Cambridge, MA 02138, USA}
\def \byu {Department of Physics and Astronomy, Brigham Young University, Provo, UT 84602, USA}
\def \apl {Applied Physics Lab, Johns Hopkins University,\\ 11100 Johns Hopkins Road, Laurel, Maryland 20723, USA}
\def \ucla {Department of Earth, Planetary and Space Sciences, University of California Los Angeles, 595 Charles E. Young Dr. East, Los Angeles, CA 90095, USA}
\def \carnegie {Earth and Planets Laboratory, Carnegie Institution for Science, Washington, DC 20015}
\def \stgallen {School of Computer Science, University of St. Gallen,\\ Rosenbergstrasse 30, CH-9000 St. Gallen, Switzerland}

\shorttitle{DEEP VI: Multi-night}
\shortauthors{Smotherman et al.}

\begin{document}
\title{The DECam Ecliptic Exploration Project (DEEP) VI: first multi-year observations of trans-Neptunian objects}
\correspondingauthor{Pedro H. Bernardinelli}
\author{Hayden Smotherman} 
\affiliation{\UW}

\author[0000-0003-0743-9422]{Pedro H. Bernardinelli}
\altaffiliation{DiRAC Postdoctoral Fellow}
\affiliation{\UW}
\email{phbern@uw.edu}

\author[0000-0001-8132-8056]{Stephen K. N. Portillo}
\affiliation{Department of Mathematical and Physical Sciences, Concordia University of Edmonton, 7128 Ada Boulevard, Edmonton, AB, T5B 4E4 Canada}
\affiliation{\UW}

\author{Andrew J. Connolly} 
\affiliation{\UW}
\author{J. Bryce Kalmbach}
 \affiliation{\UW}
\author{Steven Stetzler} 
\affiliation{\UW}

\author[0000-0003-1996-9252]{Mario Juri\'c}
\affiliation{\UW}

\author[0000-0002-1312-5529]{Dino Bekte\v{s}evi\'{c}}
\affiliation{\UW}

\author[0000-0001-7574-4440]{Zachary Langford}
\affiliation{Department of Physics and Astronomy, University of Pennsylvania, Philadelphia, PA, USA}
\affiliation{\UW}
\author[0000-0002-8167-1767]{Fred C.~Adams}
\affiliation{\UMPhysics}
\affiliation{\UMAstronomy}

\author[0000-0001-5750-4953]{William J. Oldroyd}
\affiliation{\NAU}

\author[0000-0001-8550-6788]{Matthew J. Holman}
\affiliation{\cfa}

\author[0000-0001-7335-1715]{Colin Orion Chandler}
\affiliation{\UW}
\affiliation{LSST Interdisciplinary Network for Collaboration and Computing, 933 N. Cherry Avenue, Tucson AZ 85721}
\affiliation{\NAU}

\author[0000-0002-5211-0020]{Cesar Fuentes}
\affiliation{\uchile}

\author[0000-0001-6942-2736]{David~W.~Gerdes}
\affiliation{\UMPhysics}
\affiliation{\UMAstronomy}

\author[0000-0001-7737-6784]{Hsing~Wen~Lin (\begin{CJK*}{UTF8}{gbsn} 林省文\end{CJK*})}
\affiliation{\UMPhysics}

\author[0000-0002-2486-1118]{Larissa Markwardt}
\affiliation{\UMPhysics}

\author{Andrew McNeill}
\affiliation{\NAU}
\affiliation{Department of Physics, Lehigh University, 16 Memorial Drive East, Bethlehem, PA, 18015, USA}

\author[0000-0002-7817-3388]{Michael Mommert}
\affiliation{\stgallen}

\author[0000-0003-4827-5049]{Kevin J. Napier}
\affiliation{\UMPhysics}

\author[0000-0001-5133-6303]{Matthew J. Payne}
\affiliation{\cfa}

\author[0000-0003-1080-9770]{Darin Ragozzine}
\affiliation{\byu}

\author[0000-0002-9939-9976]{Andrew S. Rivkin}
\affiliation{\apl}

\author{Hilke Schlichting}
\affiliation{\ucla}

\author[0000-0003-3145-8682]{Scott S. Sheppard}
\affiliation{\carnegie}

\author[0000-0001-6350-807X]{Ryder Strauss}
\affiliation{\NAU}

\author[0000-0003-4580-3790]{David E. Trilling}
\affiliation{\NAU}

\author[0000-0001-9859-0894]{Chadwick A. Trujillo}
\affiliation{\NAU}

\begin{abstract}

We present the first set of trans-Neptunian objects (TNOs) observed on multiple nights in data taken from the DECam Ecliptic Exploration Project (DEEP). Of these 110 TNOs, 105 do not coincide with previously known TNOs and appear to be new discoveries. Each individual detection for our objects resulted from a digital tracking search at TNO rates of motion,  using two to four hour exposure sets, and the detections were subsequently  linked  across  multiple  observing  seasons. This procedure allows us to find objects with magnitudes $m_{VR} \approx 26$. The object discovery processing also included a comprehensive population of objects injected into the images, with a recovery and linking rate of at least $94\%$. The  final  orbits  were obtained using  a  specialized  orbit  fitting  procedure  that  accounts  for  the  positional  errors  derived  from  the  digital  tracking  procedure. Our results include robust orbits and magnitudes for classical TNOs with absolute magnitudes $H \sim 10$, as well as a dynamically detached object found at 76 au (semi-major axis $a\approx 77 \, \mathrm{au}$). We find a disagreement between our population of classical TNOs and the CFEPS-L7 three component model for the Kuiper belt.
\end{abstract}

\section{Introduction} \label{sec:intro}

Beyond Neptune, there exist thousands of known bodies, called Trans-Neptunian Objects (TNOs). Due to current observational capabilities, the majority of the known TNOs have diameters of order 100 km, corresponding to the observational limits of the surveys of the last two decades \citep[see][for a review]{DEEPII}. TNOs can be separated into multiple dynamically-distinct subpopulations, including the classical Kuiper Belt, the scattering disk, resonant TNOs, and the detached population \citep{Gladman_2008}. Explaining this structure has been an active subject of research, as these small bodies are important probes of the formation and evolution of the early Solar System \citep{Nesvorny_2018_annualrev,Gladman_2021}, providing key observational constraints to models of an early giant planet instability in the Solar System \citep{NiceModel}. The migration of the giant planets can lead to objects being captured in mean motion resonances \citep{Malhotra1995} or scattered into orbits with significant eccentricities and inclinations \citep{Duncan1997}, and their present day orbital distribution are indicative of dynamical processes in the early Solar System \citep{MorbidelliAlessandro2019Kbfa}.

Since the detection of the first KBO in 1992 \citep{Jewitt1993}, astronomical surveys have consistently added to the number of known TNOs \citep[e.g.][]{Petit_2011,Sheppard2016,Bannister_2018,Bernardinelli2022}. As of June 2023, the JPL small body database\footnote{\url{https://ssd.jpl.nasa.gov/}} lists 4463 known TNOs. The majority of these objects have been discovered in surveys that detect objects in individual images, link those to detections in other images at different epochs, and use the collection of observations to determine or constrain the orbit of the discovered TNO.  The depth of this type of survey is limited by the depth of the individual exposures: the motion of the objects restricts the duration of individual exposures, as once an object has moved more than the angular extent of the image quality or ``seeing,'' trailing limits the resulting signal-to-noise ratio ($S/N$). A number of deeper surveys, typically limited to narrow fields \citep[e.g.][]{Gladman1998,Allen2001,Bernstein_2004,FRASER2008827,Fuentes2009} or relatively short time baselines \citep[e.g.][]{Smotherman_2021, kbmod}, overcome this limitation using a technique known as ``shift and stack'' or ``digital tracking''. This technique works by combining the signal from a series of short, untrailed exposures along hypothesized sky plane trajectories.  Because the orbits are not known {\it a priori}, a range of rates and directions of motion are tested.  This allows the detection of sources significantly fainter than the single image detection limit. Here, we present the first multi-year (2019-2021) linkages of TNOs detected with digital tracking in the $30\deg^2$ of the B1 fields of the DECam Ecliptic Exploration Project \citep[DEEP,][]{DEEPI}. These time spans lead to high-quality orbit determination that enable dynamical classifications \citep{Gladman_2008}. The combination of the dynamical properties with the well-understood observational biases \citep{DEEPIII} make statistical studies of our objects possible.





In Section \ref{sec:data} we briefly discuss the data processing of these fields as well as the single-night observations. In Section \ref{sec:tech}, we discuss the methods used to link these single-night observations to multi-year arcs, then fit orbits to these arcs. In Section \ref{sec:results}, we discuss the implication of our recovered TNOs and their orbital properties when compared to existing TNO distributions. In Section \ref{sec:discuss}, we summarize our results and discuss future work.

\section{Data} \label{sec:data}

As its acronym indicates, DEEP uses the Dark Energy Camera (DECam) on the 4m Blanco telescope at the Cerro Tololo Inter-American Observatory (CTIO) \citep{Flaugher_DECam_Instrument}. The data are taken in the \textit{VR} band, which has a similar band center to, for example, the Pan-STARRS $r_{p1}$ band, but broader wavelength coverage. This makes it well suited to detecting faint sources, although it is less preferred for precision photometry.

\begin{figure}[p]
    \centering
    \includegraphics[width=\textwidth]{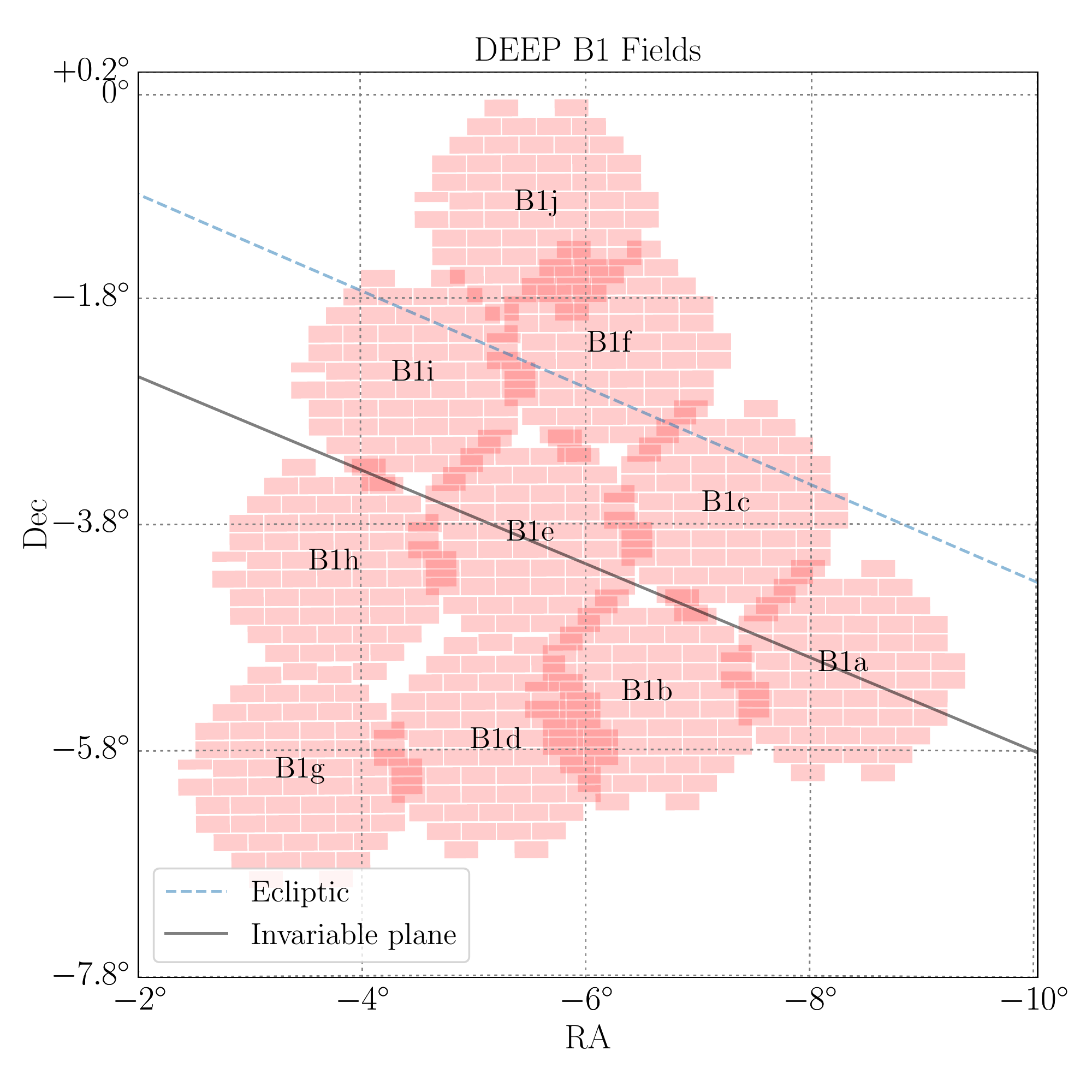}
    \caption{On-sky positions of the B1 field. The pointings included here are the first observation of each field taken in 2021. Each field is centered on the RA and Dec of the first exposure in the long stare. The solid line shows the invariable plane, and the dashed line shows the ecliptic.}
    \label{fig:DEEP_B1_Fields}
\end{figure}

The search area is separated into four quadrants, ``A0'', ``A1'', ``B0'' and ``B1''. Each quadrant is composed of a number of fields that spread out in a triangular pattern that follows the typical dispersion of on-sky motions of the trans-Neptunian over time \citep{DEEPII}. All of our fields are aligned with the invariable plane (the solid black line in Figure \ref{fig:DEEP_B1_Fields}) of the Solar System, to maximize object discovery. Each of these quadrants was covered with a series of 3 square degrees pointings that were visited over multiple nights and years. A field visit in a night consists of a ``long stare'',  a series of between 52 and 130 consecutive or nearly-consecutive 120 second exposures, which enables the use of digital tracking to discover objects. For the maximal TNO rate of motion of $5\arcsec/\mathrm{hour}$, in the typical four hours of these long stares a TNO will have moved $20\arcsec$, so the majority of objects will not move between CCDs during the stare. However, by revising these fields several times, after a few weeks or months (see Table \ref{tbl4:DEEP_Fields}), we can eventually determine the orbits of most of our objects. The overall survey time baseline means that candidates that can be linked over the entire survey will have observational arcs extending for over two years (three oppositions). 

This work focuses on the 2019, 2020, and 2021 data of the B1 quadrant, which has 10 unique fields and 29 total nights of data. Subsequent publications will analyze the remainder of the DEEP data. The total footprint of the B1 data covers about 30 square degrees and is shown in Figure \ref{fig:DEEP_B1_Fields}. The fields, nights, and number of exposures of all the data in the B1 quandrant are presented in Table \ref{tbl4:DEEP_Fields}. The mean number of exposures in this quadrant is 87, and the median is 96. For example, for field B1a on night 2019-08-28, there are 102 exposures in the long stare, with a total open shutter time of 12,240 seconds, or 3 hours 24 minutes. This long effective exposure time, combined with digital tracking, is what allows DEEP to detect such faint TNOs ($m_{VR} \approx 26.0$) from the ground.

\begin{longtable}{llrrr}
Field &      Night &         RA &       Dec &  $N_{\mathrm{exp}}$ \\
\hline
\endhead

\multicolumn{5}{r}{{Continued on next page}} \\

\endfoot

\endlastfoot
  B1a & 2019-08-28 & 351.877200 & -5.035722 &                     102 \\
  B1a & 2019-09-26 & 351.380700 & -5.240305 &                      95 \\
  B1a & 2020-10-18 & 351.121033 & -5.346222 &                      91 \\
  B1a & 2021-09-09 & 351.713325 & -5.103250 &                      65 \\
  B1b & 2019-08-29 & 353.618654 & -5.297667 &                     101 \\
  B1b & 2019-09-27 & 353.120700 & -5.502861 &                      97 \\
  B1b & 2020-10-15 & 352.863196 & -5.609639 &                     100 \\
  B1b & 2021-09-12 & 353.455079 & -5.365278 &                      70 \\
  B1b & 2021-10-01 & 353.058450 & -5.529917 &                      81 \\
  B1c & 2019-08-27 & 352.919908 & -3.623028 &                     103 \\
  B1c & 2019-09-28 & 352.424117 & -3.827972 &                      96 \\
  B1c & 2020-10-16 & 352.165621 & -3.934944 &                      96 \\
  B1c & 2021-09-06 & 352.756529 & -3.690555 &                      53 \\
  B1c & 2021-10-04 & 352.360275 & -3.854361 &                     101 \\
  B1d & 2020-10-19 & 354.606779 & -5.868861 &                      99 \\
  B1d & 2020-10-21 & 354.606525 & -5.869916 &                      99 \\
  B1d & 2021-09-27 & 354.800983 & -5.788139 &                      89 \\
  B1e & 2020-10-17 & 353.904871 & -4.195444 &                      96 \\
  B1e & 2021-09-04 & 354.495283 & -3.950028 &                      59 \\
  B1e & 2021-10-06 & 354.098908 & -4.115194 &                      97 \\
  B1f & 2020-10-20 & 353.205779 & -2.521444 &                      98 \\
  B1f & 2021-09-03 & 353.795867 & -2.276583 &                      61 \\
  B1f & 2021-10-02 & 353.400358 & -2.441222 &                      89 \\
  B1g & 2021-09-28 & 356.545621 & -6.043333 &                      92 \\
  B1h & 2021-09-08 & 356.235704 & -4.206055 &                      70 \\
  B1h & 2021-10-05 & 355.839704 & -4.371417 &                      97 \\
  B1i & 2021-09-05 & 355.531650 & -2.533694 &                      52 \\
  B1i & 2021-10-03 & 355.136983 & -2.698944 &                     100 \\
  B1j & 2021-09-30 & 354.436454 & -1.026361 &                      85 \\

\caption{Fields, nights, RA, Dec, and number of observations ($N_{\mathrm{obs}}$) for all of the nights in the DEEP B1 quadrant that are searched with \texttt{KBMOD}. RA and Dec correspond to the central RA and Dec of DECam for the first exposure in the long stare. Each exposure in each long stare is 120 seconds long.}
\label{tbl4:DEEP_Fields}
\end{longtable}

Figure \ref{fig:DEEP_B1_Fields} shows the pointing of each field in the first night that field was observed in the 2021 observing campaign. Over various nights, the pointing center for any given gield can vary by up to 0.817 degrees, with a median value of 0.39 degrees, from other observations of that field. These intra-field offsets were designed in part to mitigate the effect of chip gaps as well as the gaps between fields. Within a long stare, the median standard deviation is $\sigma_\mathrm{RA} = 0.629$ arcsec and $\sigma_\mathrm{Dec} = 0.070$ arcsec, and the images are aligned to a common WCS, as required by the shift and stack procedure.

\subsection{Data pre-processing} \label{subsec:DataPreprocessing}

We use the Legacy Survey of Space and Time (LSST) Science Pipelines to process the DEEP B1 data \citep{LSST_DM}. For these data, we start with DECam raw (uncalibrated) image files, flat field images, and bias images. Each long stare has associated flats and biases from the same night, with the exception of the 2020-10-15 visit of field B1b and the 2020-10-19 visit of field B1d, where we use flats and biases from the visits in 2020-10-16 and 2020-10-18, respectively. We then use the LSST Science Pipelines to perform instrument signal removal (ISR), photometric calibration, astrometric calibration, and image differencing. Photometric calibration is performed using Pan-STARRs DR1 data \citep{Flewelling_2020} as a reference catalog, mapping the $VR$ band to the reference $r_{p1}$ band. However, we elected to be explicit in our reported magnitudes that we are using this wide-band filter, so in all references below we have the $VR$ subscript when referring to magnitudes. 
 Astrometric calibration is performed using Gaia DR2 data \citep{gaia_dr2} as a reference catalog. We will follow the convention in the LSST Science Pipelines and refer to these photometrically and astrometrically calibrated exposures as ``\texttt{calexp}s''. At this point, cosmic rays are masked in the individual images, so that they do not introduce correlated noise in the shift and stack procedure.

After calibration, we inject a population of synthetic TNOs (or ``fakes'') into the \texttt{calexp}s, as described in detail in \cite{DEEPIII}. The goal of this population is to test all physically plausible TNO orbits, rather than attempting to reproduce any real existing TNO population structure. To accomplish this, we create a joint distribution from a low-$e$, low-$i$ Kuiper belt-like population that spans $30 < a < 80$ au; a moderate-eccentricity population with $0<e<0.4$ and inclinations uniformly spanning $0\deg$ to $90\deg$; and an isotropic distribution with randomly-sampled Cartesian coordinates between 25 and 1000 au. We propagate these objects to the epoch of each image, adding objects that intersected with the plane of a CCD to the synthetic catalog. This creates a catalog of 5737 unique synthetic objects.  These objects each overlap with a CCD image 1 to 9 times. We assign values for $m_{VR}$ from two uniform distributions, $20<m_{VR}<24$ and $24<m_{VR}<28$, with 20\% and 80\% of the total objects in the brighter and fainter distributions, respectively. As our goal is to measure our discovery efficiency as a function of magnitude, these uniform distributions ensure that we have enough sources at a given magnitude to properly reconstruct our discoverability. Note that the discovery efficiency is equivalent to the ratio of the number of detected objects to the number of injected objects, so it is independent of the shape of the underlying injected population - see \cite{DEEPIII} for a thorough discussion. 
 We propagate magnitudes with respect to the changing heliocentric and geocentric distances, such that the reference magnitude for the object coincides with the magnitude at the reference epoch for the orbital elements. Half of the synthetic population included simulated lightcurves with periods between 2 and 100 hours and amplitudes between 0 and 0.5 mag.

After injecting synthetic TNOs, we generate template images for the difference imaging procedure by coadding the long-stare images from each night. Notably for this dataset, we use the default LSST Science Pipelines coaddition statistic of \texttt{MEAN}, where the coadded image corresponds to the mean flux in each pixel that contributes to the stack. \citet{Smotherman_2021} describes in more detail the the image differencing technique, as well as the astrometric performance of the LSST Science Pipelines applied to DECam. Because we generate coadded templates from the images of each long stare, we are self-subtracting some flux from the TNOs, both real and synthetic, in the individual images. This removal will depend on the speed of the TNO and the time baseline of the specific long stare. This has the effect of reducing our $S/N$ (and thus reducing our effective depth) and complicating the magnitude measurement for each candidate object. For detected candidate objects, we model and account for this flux loss, as described in Section \ref{sec:DetectAndChar}. Future DEEP data releases will investigate the use of a non-default coaddition statistics to mitigate our flux self-subtraction loss. We then subtract the template from each \texttt{calexp} to create difference images, in which we conduct our object search. 

\section{Detection and Characterization using shift and stack} \label{sec:DetectAndChar}

We use the Kernel Based Moving Object Detection \citep[\texttt{KBMOD},][]{kbmod,Smotherman_2021} pipeline to perform digital tracking on each long stare in the data set. \texttt{KBMOD} is a GPU-accelerated digital tracking pipeline that uses likelihood images, as defined below, to estimate the likelihood that there is a source following any given trajectory, represented by a set of velocities in which the images are stacked, using all pixels in the first image as an origin point. Our grid choice for the \texttt{KBMOD} trajectories as applied to DEEP is as follows: velocity ranges from 130 pixels/day to 400 pixels/day (1.4 arcsec/hr to 4.5 arcsec/hr, for comparison, at 40 au, a TNO's typical range of motion is 3 arcsec/hr) in 50 uniform steps.  Angles are $\pm45^\circ$ of the direction of decreasing ecliptic longitude, divided uniformly into 30 steps, so that the maximum separation between neighboring trajectories would be less than about 2 PSF FWHM over a 4 hour time baseline. This grid choice exhausts the entirety of the Kuiper belt, and the majority of bound orbits between 30 and 80 au \citep[see][for a detailed presentation]{DEEPIII}.

We define
\begin{align}\label{eqn:PsiPhi}
    \Psi(y) =& \sum_j \frac{1}{\sigma_j^2}n(x_j)T(x_j-y)\\
    \Phi(y) =& \sum_j \frac{1}{\sigma_j^2}T(x_j-y)^2,
\end{align}
where $\sigma_j$ is the variance at pixel $j$, $n(x_j)$ is the number of counts in pixel $j$ at position $x_j$ in a given difference image, and $T(x_j-y)$ is the point spread function (PSF) centered at the true position $y$ of a given point source. This means that $\Psi$, the likelihood image, represents the cross-correlation of the PSF and the data, a quantity equivalent to the convolution of the transpose of the PSF with the data (for a symmetric PSF, this reduces to the convolution of the PSF and the data). $\Psi$ is simply the effective area of the PSF when weighted by the inverse variance of the counts in each pixel. These quantities lead to an optimal coaddition for point sources \citep{kbmod}. In this framework, sources in the image represented by $\nu = \Psi/\sqrt{\Psi}$ are $\chi^2$ distributed, and so, sources above a certain threshold $\nu_\mathrm{thres}$ are detections of this significance. 

When coadding along images $i$ in a trajectory, we have that:
\begin{align}
    \Psi_\mathrm{coadd} =& \sum_i \Psi_i\label{eqn:PsiCoadd}\\
    \Phi_\mathrm{coadd} =& \sum_i \Phi_i\label{eqn:PhiCoadd}\\
    \nu_\mathrm{coadd} =& \Psi_\mathrm{coadd}/\sqrt{\Phi_\mathrm{coadd}},\label{eq:sumlike}
\end{align} where $\nu$ is the $S/N$ of a source in the coadd. Following the notation of \cite{Smotherman_2021}, we denote by $\sum LH$ the summed likelihood (equivalent to the $S/N$ in the pre-convolved images) of a single trajectory. In other words, the summed likelihood is the $S/N$ as given by equation \ref{eq:sumlike} for a initial pixel $(x,y)$ and trajectory. We refer the reader to \citet{kbmod} for more detailed description of the \texttt{KBMOD} procedure.

This formalism allows the core GPU algorithm in \texttt{KBMOD} to evaluate more than $10^{10}$ candidate trajectories in a few minutes using consumer-grade GPUs \citep{kbmod}. For DEEP, a \texttt{KBMOD} search of a stack of $\mathcal{O}(100)$ images for a single CCD has a median runtime of 576s using an NVIDIA A40 GPU. This includes loading the data from disk, running the core GPU algorithm, filtering false positives and generating image stamps. Our total runtime for this search was about 282 hours.  

In order to characterize our false positive rate, we also run ``reverse'' searches in the direction of increasing ecliptic longitude, which is opposite from the direction of a TNO due to the reflex motion of the Earth. These reverse searches will never yield a real TNO detection, but will lead to false positive detections due to the (trajectory independent) correlated noise (e.g. difference imaging artifacts and poorly subtracted sources) in the image series, and thus represent an accurate characterization of the false positive rate as a function of $\sum LH$. We set the minimum sum likelihood threshold for a candidate trajectory to be considered to $\sum{LH}>7$. This limit was we defined by studying the number of candidates yielded in the reverse searches (where we expect that the majority of detections are false positives) vs $\sum LH$: below $\sum LH = 7$, the number of detections coming from the reverse searches increases by an order of magnitude.

Due to the number of false positives at very slow speeds (130 to 150 pixels/day, where the effect of the correlated noise sources is more pronounced) or in dense CCDs where the images were poorly differenced (when the number of trajectories with $\sum{LH} > 7$ at velocities larger than 150 pixel/day\footnote{This ``overflow'' occurred in 51/1680 of the forward searches.}), we increase the likelihood threshold from $\sum{LH}>7$ to $\sum{LH}>10$. This empirically determined limit significantly reduces the number of false positive trajectories that result from the search.

\begin{figure}[h]
    \centering
    \includegraphics[width=1\textwidth]{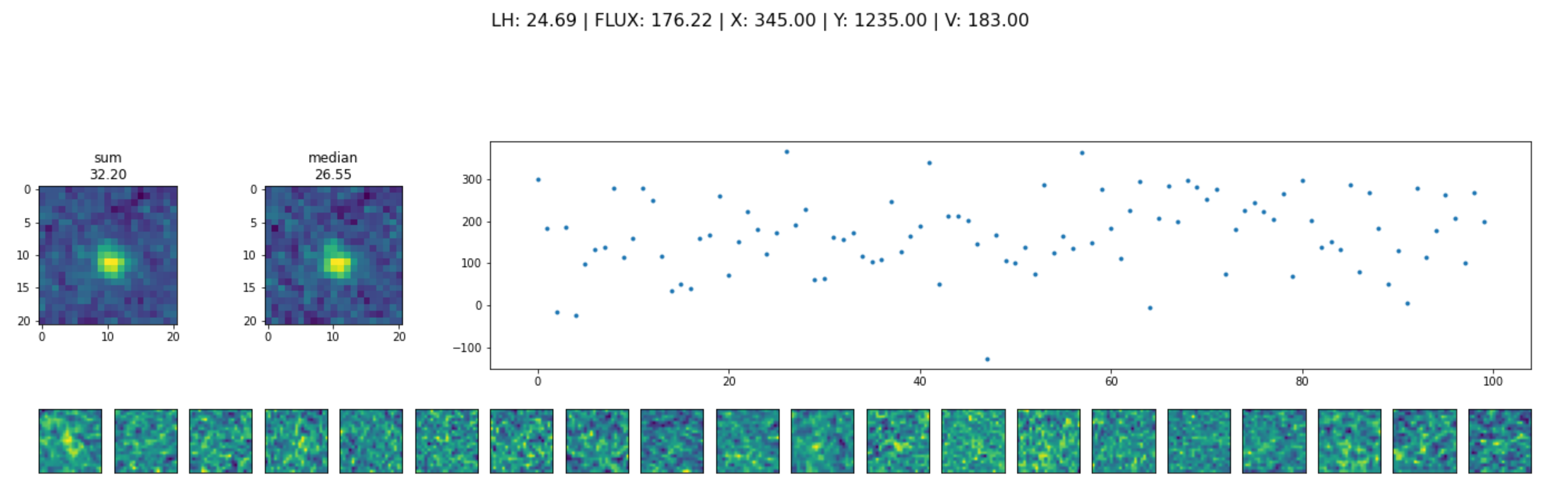}
    \includegraphics[width=1\textwidth]{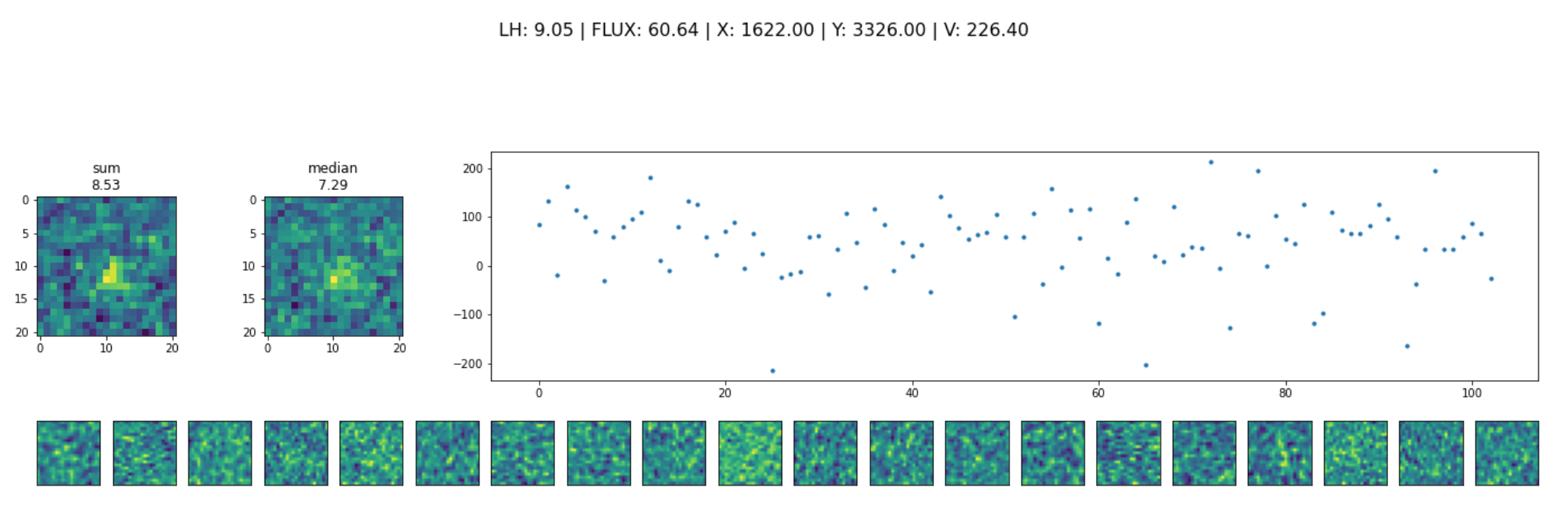}
    \caption{Two examples of objects recovered with \texttt{KBMOD}, the first an injected TNO with magnitude $m_{VR}=24.9$ and a barycentric distance of 60 au, the second a real cold Classical TNO with $m_{VR} = 26.0$. These images are used for human vetting. The title includes information from the \texttt{KBMOD} search, including $\sum{LH}$, flux, starting x and y pixel location, and speed. The two postage stamps in the upper left are sum and median coadds, as well as an approximate $S/N$ estimated from these postage stamps. The graph in the upper right shows an approximate flux lightcurve. The postage stamps included below are ther first 20 our of 100 cutouts from individual images in the long stare.}
    \label{fig:KBMOD_Candidate}
\end{figure}

In order to further reduce the number of false positive trajectories in the search, we apply the same convolutional neural network (CNN) from \citet{Smotherman_2021} on the resulting coadded stamps from the trajectories. This CNN discards candidates with a CNN-assigned probability of being from a ``good'' source of less than 50\%. We note that in practice this threshold is merely a user-defined cutoff, rather than any kind of robust statistical likelihood - this choice ensures a good trade-off between true positives/false negatives and false positives. For the candidate trajectories that remain after the aforementioned cuts, we assign a randomized candidate id and generate images for human vetting. This is done with the combined results from both the forward and the reverse searches, so that we can estimate the false negative and false positive rate of the vetting procedure, as shown below. We show two examples of the images used for human vetting in Figure \ref{fig:KBMOD_Candidate}. This procedure helps further identify artifacts such as poorly substracted sources that otherwise produced an acceptable, PSF-like stack, as well as identify other cases where the CNN does not identify a false positive (\emph{e.g.} unmasked diffraction spikes). This procedure rejected about half of the candidates coming from the CNN filtering.

After human vetting, we further improve the quality of the determined trajectory and to characterize our errors in the starting and ending position, flux, and magnitude. We perform this analysis on all candidates that pass human vetting. This method uses the PSF determined for each image, as opposed to a constant-width gaussian PSF as used by \texttt{KBMOD}. The process for applying this ``joint-fit'' characterization is as follows. First, we load 31x31 pixel warped difference imaging and \texttt{calexp} cutouts, PSFs, magnitude zero points, stamp centers in R.A./dec and pixel space, and associated world coordinate systems (WCSs). Next, we rescale the stamps to a common magnitude zero point of $m_{VR,0}=31$. After loading the data, we run a high-precision positional fit. We use the same formalism as \texttt{KBMOD} for flux and likelihood estimation, and minimize $\sum LH$ to find the topocentric best-fit trajectory, that is, the statistically optimal four dimensional solution of the starting and ending positions, which are each bounded to be no more than 10 pixels from the center of the corresponding stamp. By computing the $4\times4$ Hessian matrix of this $\sum LH$ minimization, we also determine the (astrometric) covariance matrix (defined as the inverse of the Hessian) of the starting and ending positions. These will be incorporated in our orbit fitting procedure in Section \ref{sec:tracklet}.

Finally, we measure the magnitude of each candidate object. When fitting magnitudes, we adjust the PSF of each image to model the effect of the ``negative well'' in the difference images. This ``negative well'' is caused by flux that was self-subtracted in the coaddition procedure. We model this effect by subtracting the relevant PSF scaled by the inverse of number of images at the location the candidate object for each image in the stack. Using this approach, we first measure the flux in each individual image in a single candidate trajectory. We use sigma-clippping to iteratively reject 3-$\sigma$ flux outliers, and, for the cutouts that are not clipped, we then measure the variance-weighted flux and flux uncertainty using the same formalism as \texttt{KBMOD}. Next, we add an additional flux term in quadrature to the flux uncertainties such that the sigma-clipped standard deviation of the distribution of the residuals of the implanted and measured fluxes for the simulated objects corresponds to a unit Gaussian distribution. This helps to ensure we do not underestimate our uncertainties in flux. Future processings of the DEEP data will mitigate this problem by employing more refined flux measurement techniques in the presence of background sources \citep[see][for a recent example]{bernardinelli2023}.

\section{Methodology} \label{sec:tech}

\subsection{$S/N$ clipping of candidate single-night detections}
We refer to candidates that are potential moving objects (as opposed to implanted fakes or false positives) as ``real''. First, to further ensure no fakes are in this sample, we run fake association on both starting and ending RA and Dec. From our list of candidate objects that pass human vetting, we remove any candidate object that is within 5 arcsec of the starting position or the ending position of a fake in the implanted fakes catalog. Only six additional candidates are associated with fakes when using either starting or ending position (rather than solely starting position). Regardless, we exclude these six additional candidates from the real distribution in the following analysis.

First, we again run the joint-fit, this time on the reverse search candidates that also passed human vetting. We then use these joint-fit results to assign a signal-to-noise ratio to each detection. We use this second run of the joint-fit to identify an appropriate $S/N$ cutoff. Here, $S/N$ is functionally identical to $\sum LH$, except it uses the $\Phi$ and $\Psi$ values from the joint-fit and is therefore more closely aligned to the actual $S/N$ of the source. Namely, $S/N = \sum_i \Psi_i / \sum_i \sqrt{\Phi_i}$, where $\Psi_i$ and $\Phi_i$ are generated using the negative well modelling described in Section \ref{sec:DetectAndChar}.

\begin{figure}[tbh]
    \centering
    \includegraphics[width=1\textwidth]{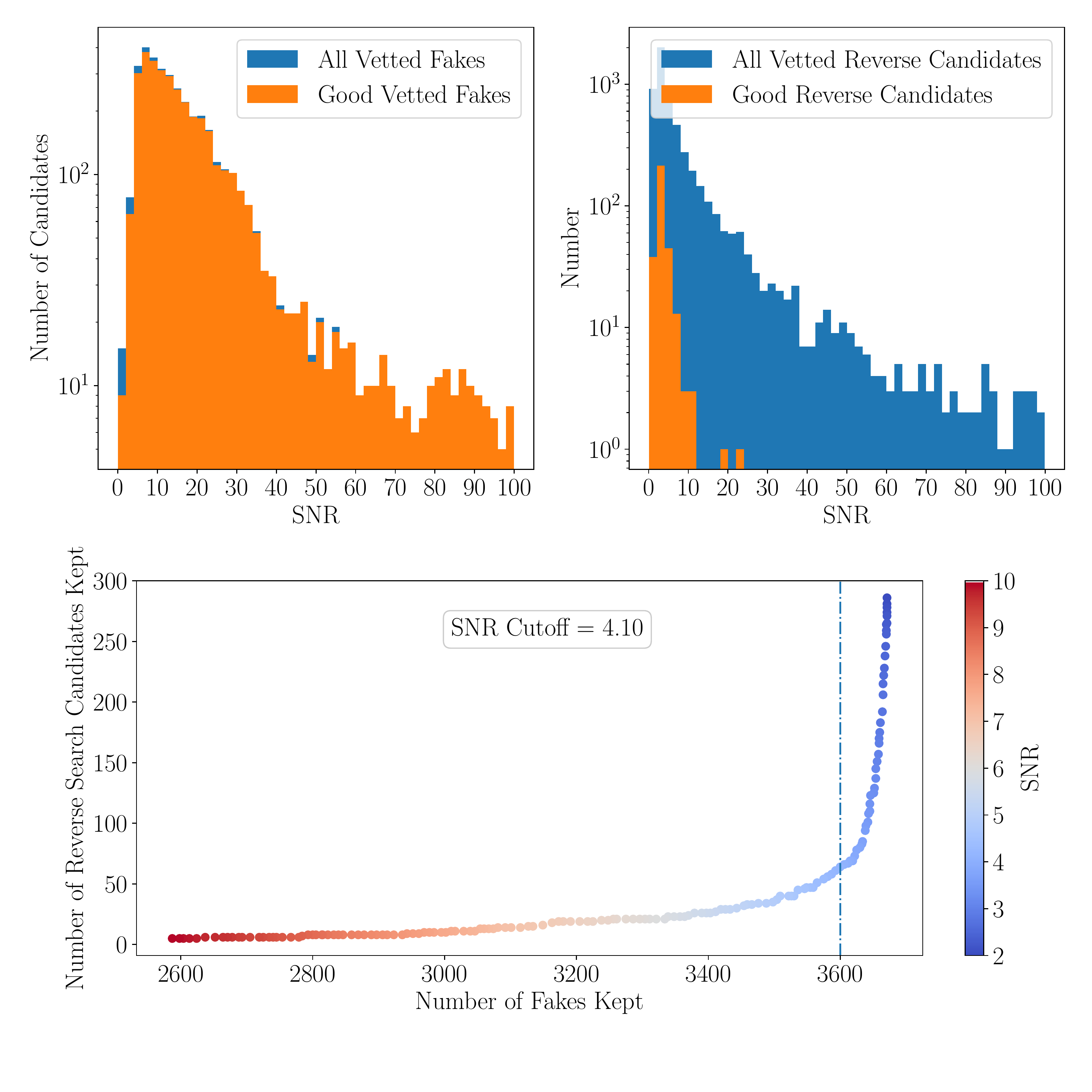}
    \caption{A visualization of the signal-to-noise ratio cutoff used to determine if a candidate should be considered real. \textit{Top left:} The number of all fakes that were included in vetting (blue) and the number of fakes that were labeled ``good'' in human vetting. The ideal case is that the orange histogram matches the blue histogram. $S/N$ is bounded to $0<S/N<100$ for visibility. \textit{Top right:} The number of all reverse search candidates that were included in vetting (blue) and the number of reverse search candidates that were labeled ``good'' in human vetting (orange). The ideal case is that the orange histogram goes to zero. We bound to $0<S/N<100$ for visibility. \textit{Bottom:} A parametric plot of the number of reverse search candidates kept after human vetting (lower is better) versus the number of fakes kept after human vetting (higher is better) as a function of $S/N$. The blue dash-dot vertical line corresponds to the selected $S/N$ cutoff value of 4.10.}
    \label{fig:DEEP_$S/N$_Cutoff}
\end{figure}

In order to find the ideal $S/N$ cutoff, above which we can classify a non-fake forward search candidate as real, we use our recovered fake distribution and our reverse search distribution. To characterize our false positive rate, we ran reverse searches on all stacks of CCD exposures in the long stares. This resulted in a total of 6973 reverse search candidates included in human vetting. Of these, we erroneously labeled 350 as good. Similarly, to characterize our vetting efficiency, we compare the distribution of fakes before and after human vetting. These distributions are shown in Figure \ref{fig:DEEP_$S/N$_Cutoff} as a function of $S/N$.

Our goal is to find a $S/N$ cutoff that keeps the greatest number of (forward search) fakes while rejecting the greatest number of reverse search candidates. We quantify this cutoff by iteratively increasing the $S/N$ cutoff until the fractional decrease in the number of fakes kept is consistently greater than the fractional decrease in the number of reverse search candidates kept. Note that, as the samples of each group (real and false positives from the forward and reverse searches) are imbalanced, this cutoff does not apply equally to the two samples. Here, we define ``consistently'' to mean that the aforementioned comparison happens three times in a row. Above this cutoff, increasing the $S/N$ threshold discards more fakes than it discards reverse search candidates. For the purpose of finding the $S/N$ cutoff, we consider the range of $S/N$ from $2<S/N<10$, with an $S/N$ step size of 0.05. As shown in Figure \ref{fig:DEEP_$S/N$_Cutoff}, we find this cutoff point to be $S/N=4.10$. We will use this cutoff as one of several cuts in linking and orbit fitting.

\subsection{Astrometric Uncertainty Characterization}
\label{sec:astrometry}
\begin{figure}[tbh]
    \centering
    \includegraphics[width=\textwidth]{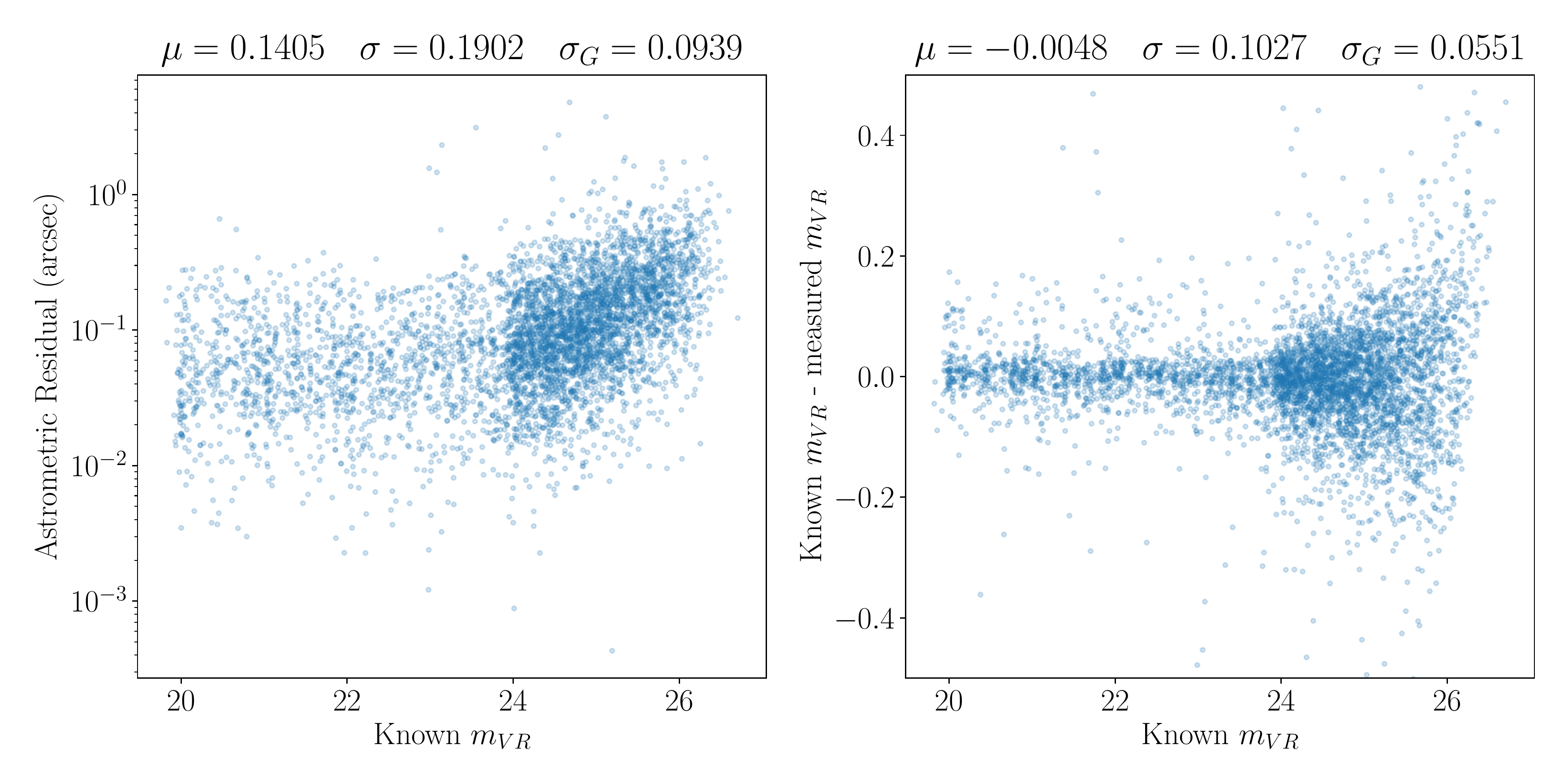}
    \caption{Astrometric (left) and photometric (right) residuals for single-night detected fakes with $S/N>4.1$. These results come from the output of the joint-fit process described in Section \ref{sec:DetectAndChar}. The title of each subplot shows the mean $\mu$, standard deviation $\sigma$, and $\sigma_G$ (see Equation \ref{eq:sigmaG}) value of each distribution.}
    \label{fig:astrom_photom}
\end{figure}

In Figure \ref{fig:astrom_photom}, we show the astrometric residuals and the photometric residuals for the joint-fit results with $S/N>4.1$. We study the total astrometric residuals
\begin{equation}
    d_\mathrm{astrom} = \sqrt{ \left( d_\alpha \cos{\delta_\mathrm{Known}}\right) ^2+d_\delta^2},
\end{equation}
where $\alpha$ corresponds to RA at the start of the long stare, $\delta$ corresponds to Dec at the start of the long stare, $d_\alpha$ is the residual in RA and $d_\delta$  the residual in Dec and, for simplicity, this measurement ignores the directional dependence of the errors, that is, the direction of the $(d_\alpha \cos \delta, d_\delta)$ vector. In addition to the mean and standard deviation, we also compute $\sigma_G$, an estimator for the underlying standard deviation of a Gaussian distribution in the presence of outliers. This is a more robust estimator of the ``characteristic'' scatter one might expect when selecting a random member of a data set \citep{astroMLText,Smotherman_2021}. Given two quantile values from a data set, $q_i$ and $q_j$, we have that
\begin{equation}
    \sigma_G\equiv C_{i,j}\left(q_j-q_i\right), \label{eq:sigmaG}
\end{equation}
where $C$ depends on the choice of quantiles. While the choice of quantiles is rather arbitrary, the 25\% and 75\% quantiles (so $C_{25,75} \approx 0.7413$) offers a good balance between a large enough number of samples from the distribution without risking significant contamination from the outliers.

The mean of the astrometric residuals is $\mu=140.5$ mas, the standard deviation is $\sigma=190.2$ mas and $\sigma_G=93.9$ mas. Given that the distribution of astrometric residuals is bounded to be positive, these are overestimating the astrometric quality of our data. We revisit this concept in Section \ref{sec:tracklet}. For the photometric residuals, the mean is $\mu=-0.0048$ mag, the standard deviation is $\sigma=0.1027$ mag, with $\sigma_G=0.0551$ mag. The standard deviation implies that our typical source has $S/N \approx 10$ ($S/N \approx 18$ if using $\sigma_G$), a reasonable result considering that we are limiting our detections to sources with $S/N \gtrsim 7$ and that our fakes are uniformly distributed in magnitude space.  

\subsection{Linking and Orbit Fitting Procedures}

Now that we have validated single-night recoveries by comparing to our population of fakes detected in a single-night long stare, we can link our single-night observations to fit detailed orbits. Furthermore, we can analyze the population of linked fakes to validate the real objects and real linked orbits that we detect. 

We parameterize each single-night detection as a tracklet of two observations defined by the starting and ending on-sky positions, and, for simplicity, we take the largest diagonal value of the covariance matrix for each set of positions. Tracklets for which the joint-fit failed to return a covariance matrix (about 10\% of the total) due to numerical instabilities in calculation of the Hessian matrix of the likelihood are given a default uncertainty in starting and ending RA and Dec of 270 mas (approximately 1 pixel). Note that, in Section \ref{sec:tracklet}, we revisit these uncertainties, so the overestimate here is not an issue.

We apply the linking methodology of \cite{Bernardinelli2019,Bernardinelli2022}\footnote{\url{https://github.com/bernardinelli/deslinker}}. We refer the reader to these papers for detailed presentations of the methods. The application of this technique to the temporally sparse Dark Energy Survey (DES) wide field data has proved to be $>99\%$ efficient in an exhaustive population of synthetic objects subject to the DES geometric selection functions and the magnitude efficiency of individual exposures \citep{Bernardinelli2022}. The initial linking uses only the information of a single position per night and proceeds in bins of inverse geocentric distance $\gamma \equiv 1/d$. We limit our search to $20 < d < 2500$ au, exhausting the range where we are able to recover TNOs in the digital tracking procedure, and enabling us to determine potential mislinkages at distances where we do \emph{not} expect to see detections. 

The process proceeds by clustering individual detections (represented by either end of a tracklet) in pairs in future observing nights, where the clustering radius defined by the limiting motion of a bound orbit at the chosen $\gamma$ for the time span. In other words, given an individual detection and $\gamma$, detections from other exposures that lie inside a circular area with this limiting radius are clustered into pairs. Triplets are found by decomposing the motion of the pair into the parallactic and binding axes defined in \cite{Bernardinelli2022}, representing deviations from the nominal search $\gamma$ and by the limiting line of sight velocity given the first two detections. We proceed by fitting the orbits of all triplets in the distance bin, with a Gaussian prior on $\gamma$. Unlike the DES application of this methodology, we find pairs and triplets of detections across the multiple observing seasons of the survey.

The major difference between the two data sets that enables this generalization is that the DEEP candidate list has a low transient density and high purity, that is, most detections are real objects, while the DES transient catalog is dominated by asteroids and image artifacts \citep{Bernardinelli2019}. Each typical DEEP pointing has $\mathcal{O}(100)$ detections, with a typical false positive rate less than one false positive per real source, while the DES search had $\mathcal{O}(10^3)$ detections per exposure, with $\mathcal{O}(1)$ real TNOs per image. 

After further detections are found in the other nights of data, we fit the candidate's orbit using the orbit fitting routines of \cite{Bernstein2000}, part of the \texttt{orbitspp}\footnote{\url{https://github.com/gbernstein/orbitspp}} package. Each candidate orbit is propagated to all other exposure epochs, and we search for further detections within the projected on-sky $5\sigma$ error ellipse. Once an additional single night candidate is found, we iterate this process with the updated orbit, until all possible combinations have been exhausted. This leads us to an initial $\chi^2$ with $\nu = 2 n- 6$ degrees of freedom ($n$ being the number of detections in the fit).

Once we have fit candidate orbits for our single-night detections, we apply cuts to all possible candidates in order to create our set of trustworthy multi-night linked orbits. We require  that the orbital arcs span $>0.8$ yr, and that the shortest arc remaining once removing either the first or the last night of the data to span $>0.5$ yr.  We also require that the object be detected in 4 or more nights. These choices were optimized so that the majority of our fakes are accepted by the linking process, while still minimizing the potential for false positives.

We require each real candidate orbit to be composed solely of real single-night detections and each fake candidate orbit to be composed solely of fake (implanted synthetic) single-night detections. We label candidate orbits with a mix of real and fake component tracklets as ``bad'', because no correct linkage can be composed of both real and fake component tracklets.

Next, we iteratively remove possible duplicate orbits. We iterate through the list of all real orbits, fake orbits, and bad orbits separately. For each orbit in the list, we find any other orbits that share at least one tracklet with the orbit. From this group, we save the orbit with the lowest $\chi^2/\nu$ and discard any other orbits that share tracklets with this orbit. If there are any remaining orbits, we again check if there are any shared tracklets and repeat the process if so. Then, we remove any candidate orbits in the bad population if that candidate orbit shares any single-night component tracklets with an orbit in the real or fake orbit populations. This process requires that each single-night long stare tracklet corresponds to only one orbit in any of the three (real, fake, or bad) populations.

At this point, we investigate which of our fakes are properly linked and which are mislinked. A mislinked fake orbit occurs when single-night tracklets from two different implanted fakes link to a single orbit, or when a fake connects to a real tracklet (that potentially belongs to a real TNO). Out of 250 candidate orbits, 20 are mislinked fakes and 230 are properly-linked fakes. We add these 20 mislinked fakes to the ``bad'' population. With these populations in mind, we can identify one final cut before investigating the real candidate orbits.

Given that we have recovered 230 of the total 244 possible candidates, we have a linking efficiency of 94\%. We investigate these 14 synthetic objects that were not recovered in the linking process. Of these, ten had significant ($> 2\arcsec$) astrometric residuals and, thus, are outliers of our residual distribution. Another two had poorly constrained positional uncertainties, with at least one sky coordinate having $\sigma \approx 1 \arcsec$, and so were rejected by the linking due to uncertain orbit fits. The final two objects did not have any obvious reason to be rejected by the linking algorithm, and so we believe those were true missed linkages. Accounting for these factors, the ``true'' linking efficiency is closer to the 99\% reported by \cite{Bernardinelli2022}, but any statistical characterization of our recovered objects must use the smaller 94\% figure (plus any cut derived from the poor quality orbit fits of Section \ref{sec:tracklet}), as these problems also affect real TNOs.

\subsection{Tracklet fitting}
\label{sec:tracklet}
As we demonstrated in Section \ref{sec:astrometry}, our shift and stack ``tracklets'' have strong correlations not only between each pixel coordinate for any given time, but also between the positions at the beginning and the end of the long stares. If we do not account for these correlations, for example, by using a circular error matrix where the radius is the semi-major axis of the error ellipse or only include the correlations between each component in a given exposure, this would make our orbit fits not properly reflect our astrometric capabilities. This would potentially lead to uncertain or incorrect dynamical classifications, and it also makes any ``fit quality'' tests (i.e. selecting on $\chi^2/\nu$) complicated, as we do not know if we can trust our uncertainties. 

To address these problems, we expand upon the methodology of \cite{Bernstein2000} to include the full correlation matrix obtained by the joint-fit. Our goal is to fit the \cite{Bernstein2000} parameter set $\hat\alpha \equiv \{ \alpha \equiv x/z, \beta \equiv y/z, \gamma \equiv 1/z, \dot\alpha \equiv \dot{x}/z, \dot\beta \equiv \dot{y}/z, \dot{\gamma} \equiv \dot{z}/z\} $, where the $z$ coordinate corresponds to the line-of-sight distance $d$ between target and observer. 

We transform the starting and ending sky positions to tangent plane coordinates $\{\theta_x,\theta_y, \theta_x',\theta_y'\}$. The difference between each tracklet for long stare $\mu$ with starting observing time $t_\mu$ and ending time $t'_\mu$, and the model coordinates $\{\tilde\theta_x(t|\hat\alpha),\tilde\theta_y(t|\hat\alpha)\}$ is

\begin{equation} 
    \Delta \boldsymbol\theta_\mu = \begin{pmatrix} \theta_{x,\mu} \\ \theta_{y,\mu} \\ \theta'_{x,\mu}\\ \theta'_{y,\mu}\end{pmatrix} - \begin{pmatrix} \tilde\theta_x (t_{\mu}| \hat\alpha) \\ \tilde\theta_y (t_{\mu} |\hat\alpha) \\ \tilde\theta_x(t'_{\mu}|\hat\alpha) \\ \tilde\theta_y(t'_{\mu}|\hat\alpha)
    
    \end{pmatrix}
\end{equation}

Thus, our goal is to find $\hat\alpha$ that minimizes
\begin{equation}
     \chi^2 = \sum_\mu \Delta \boldsymbol \theta_\mu C^{-1}_\mu \Delta \boldsymbol\theta_\mu^\top,\label{eq:chi2}
\end{equation}
where $C_\mu$ is the covariance matrix of the tracklet. We add the covariance determined by the joint fit procedure to the identity matrix scaled by the excess astrometric uncertainty determined in Section \ref{sec:astrometry}, $\sigma_\mathrm{ast}^2 = (93.9 \, \mathrm{mas}/\sqrt{2})^2$. Note that the $\sqrt{2}$ is needed as this value was determined for both RA and Dec simultaneously, while here we need an error per coordinate.

In order to maintain the numerical stability of the fit, we first fit the orbit with $\dot\gamma = 0$ and circular errors corresponding to the semi-major axis of each error ellipse. After this fit converges, we re-fit the full orbit with the $\chi^2$ in Equation \ref{eq:chi2} using the same routines as \cite{Bernstein2000} and modifications presented in \cite{Bernardinelli2019}, namely the weak bound orbit prior in their Equation (15). We do \emph{not} include their prior on $\gamma$, as with $4+$ temporally sparse tracklets the solution is well-behaved and can reach its global minimum easily. The covariance matrices for $\hat\alpha$ are calculated using the analytical derivatives of the model, as in \cite{Bernstein2000}. The reported state vectors are ICRS aligned with the Solar System barycenter as the origin, with a common epoch for all objects. In addition to the Sun, the four giant planets are treated as active particles for the orbit fit, and their positions are derived from the DE-440 ephemerides \citep{Park2021}. The derived orbital elements are barycentric and use the total mass of the eight planetary systems in addition to the Solar mass for the conversions from state vectors. 

\begin{figure}[tbh]
    \centering
    \includegraphics[width=0.49\textwidth]{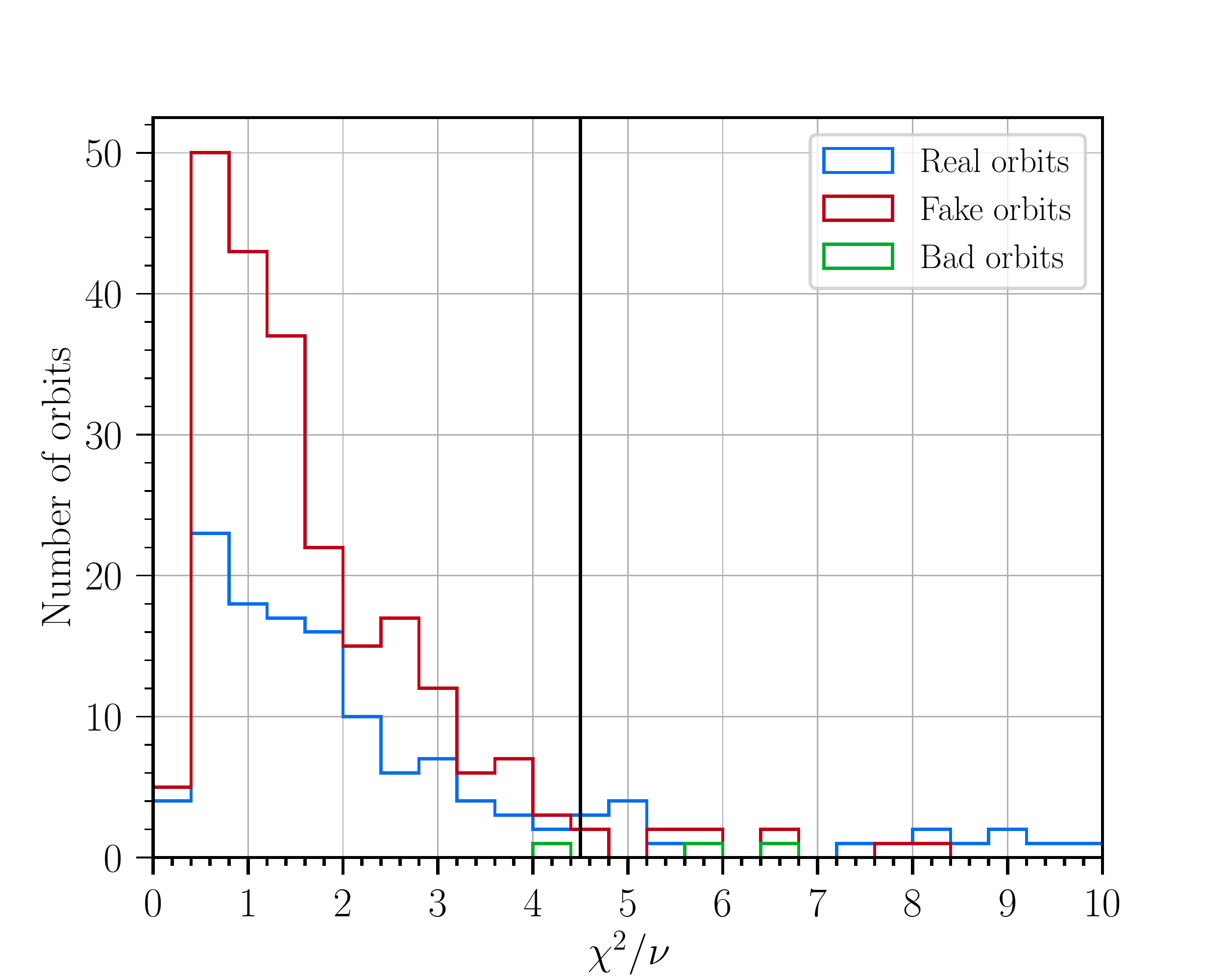}
    \includegraphics[width=0.49\textwidth]{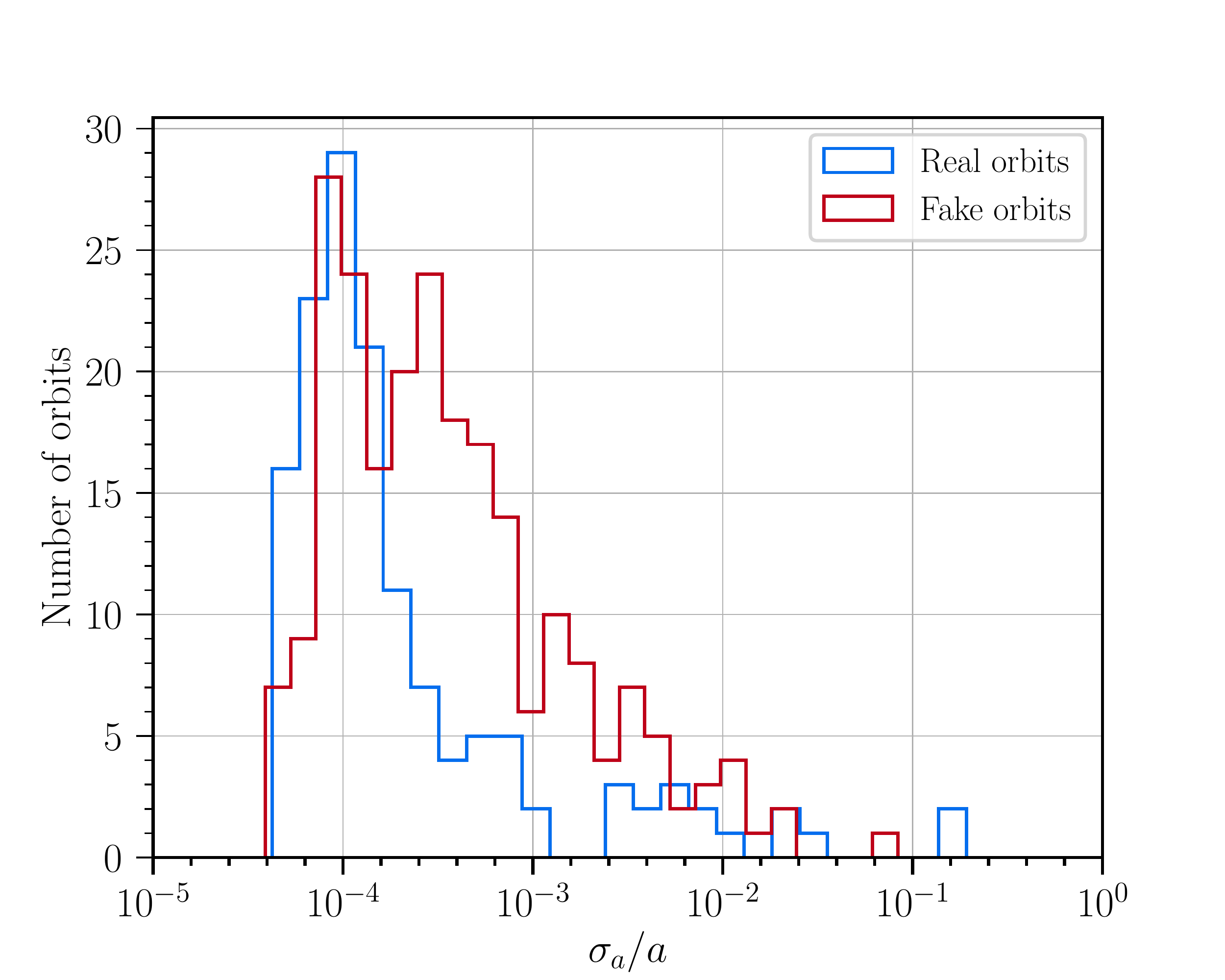}
    \caption{\emph{Left:} Histograms of $\chi^2/\nu$ determined with the full covariance matrices for the subsets of linkages corresponding to real detections (blue), fake detections (red), and the bad linkages, that is, those that we know are mislinked (orange). The vertical line indicates the $\chi^2/\nu = 4.5$ threshold. \emph{Right:} Histogram of the fractional error in semi-major axis ($\sigma_a/a$) for the real (blue) and fake (red) orbits with $\chi^2/\nu < 4.5$. We note that the different structure in the distributions is not an inherent problem, but stems from the very different populations of orbits that belong to these two samples.}
    \label{im:chi2}
\end{figure}

Now, as we've determined a new, more reliable $\chi^2/\nu$ for our objects, we can apply one additional cut to select our high-confidence orbits. Figure \ref{im:chi2} shows the distributions for the our three subsets of linkages (real, fake and bad linkages). We choose a threshold that accepts 95\% of the correctly linked fakes, $\chi^2/\nu < 4.5$, leaving us with $218/244 = 89.3\%$ succesful linkages. Of our 38 bad linkages, only 3 have $\chi^2/\nu \leq 10$ (4.1, 5.9 and 6.6, from smallest to largest), so our contamination from real mislinkages, if it exists, is minimal. Only two of our real objects have $4.0 \leq \chi^2/\nu \leq 4.5$ and, beyond this threshold, the distribution increases slightly (at 5 objects with $\chi^2/\nu \approx 5$) before dropping again.

We also note that our $\chi^2/\nu$ distributions (for both real and fake objects) peak very close to $\chi^2/\nu = 1$, indicating that the joint-fit procedure, with this additional 93 mas scatter per coordinate pair, is fully capturing our astrometric precision. The right panel of Figure \ref{im:chi2} shows the fractional error in semi-major axis $a$ for both the real and fake linkages. A typical orbit of a real object has $\sigma_a /a \approx 10^{-4}$, an orbit quality that is comparable to more ``traditional'' surveys like OSSOS \citep{Bannister_2018} and DES \citep{Bernardinelli2022} that discover objects in individual images. We note that the differences in both distributions are direct effects of the differences between the magnitude and orbital distributions of the real and fake populations: the uniform magnitude distribution of the fakes leads to a higher fraction of bright, better measured objects when compared to the approximately power law distribution of the real objects (on average fainter, thus with worse astrometric measurements). However, as the real objects (as will be seen below) are primarily classical KBOs with relatively low eccentricities, their semi-major axes are better constrained than the more eccentric population of fake objects, which reproduces the well known correlation between $a$ and $e$ in trans-Neptunian orbit recovery \citep[see, \emph{e.g.}][]{Bernstein2000}.

\section{Results}\label{sec:results}

\subsection{Object sample}

After applying the aforementioned cuts, we have a catalog of 110 detected real objects with multi-year linkages. We use \texttt{skybot} \citep{Skybot} to identify if there are known objects in the DEEP B1 data. We associate all single-night candidates input into linking with KBOs reported with \texttt{skybot}. If both the starting RA and the starting Dec of the detection are within 20 arcsec of the RA and Dec predicted by \texttt{skybot} in multiple DEEP pointings, then we consider a candidate associated with a known KBO. This leaves us with 4 known KBOs (2003 QL$_{91}$, 2002 PV$_{170}$, 2002 PX$_{170}$, and 1999 RX$_{215}$), all of which were recovered, further increasing our confidence in our search. The orbital elements, magnitudes, dynamical classifications and MPC identifiers for all of our 110 objects are presented in Table \ref{tb:objects}.

\begin{deluxetable}{cc}
  \tablecaption{Orbital properties of the TNOs from DEEP B1\label{tb:objects}}
  \tablehead{\colhead{Column name and unit} & \colhead{Description}}
  \startdata
  \texttt{MPC}  & Minor Planet Center object designation \\
  \texttt{a} (au) & Semi-major axis \\
  \texttt{e}  & Eccentricity \\
  \texttt{i} ($\deg$) & Inclination \\
  \texttt{aop} ($\deg$) & Argument of perihelion \\
  \texttt{lan} ($\deg$) & Longitude of ascending node \\
  \texttt{T\_p} (Julian years) & Time of perihelion passage  \\ 
  \texttt{sigma\_a} (au) & Uncertainty in $a$ \\
  \texttt{sigma\_e} & Uncertainty in $e$ \\
  \texttt{sigma\_i} ($\deg$) & Uncertainty in $i$ \\ 
  \texttt{sigma\_aop} ($\deg$) & Uncertainty in $\omega$ \\
  \texttt{sigma\_lan} ($\deg$) & Uncertainty in $\Omega$ \\
  \texttt{sigma\_Tp} (yr) & Uncertainty in $T_p$\\
  \texttt{m\_VR} (mag) & Mean magnitude \\
  \texttt{H\_VR} (mag) & Phase-corrected absolute magnitude  \\
  \texttt{sigma\_m} (mag) & Magnitude uncertainty (applies to $m$ and $H$) \\
  \texttt{Classification} & Dynamical classification  \\
  \texttt{Secure} & Indicates whether the dynamical classification is secure \\
  \enddata 
  \tablecomments{The table is provided in its entirety in machine-readable format (FITS), here we are only describing the columns included in the data release. All elements correspond to Julian epoch 2019-08-30, and are barycentric and ecliptic-aligned.}
\end{deluxetable}

\begin{figure}[t!]
    \centering
    \includegraphics[width=0.7\textwidth]{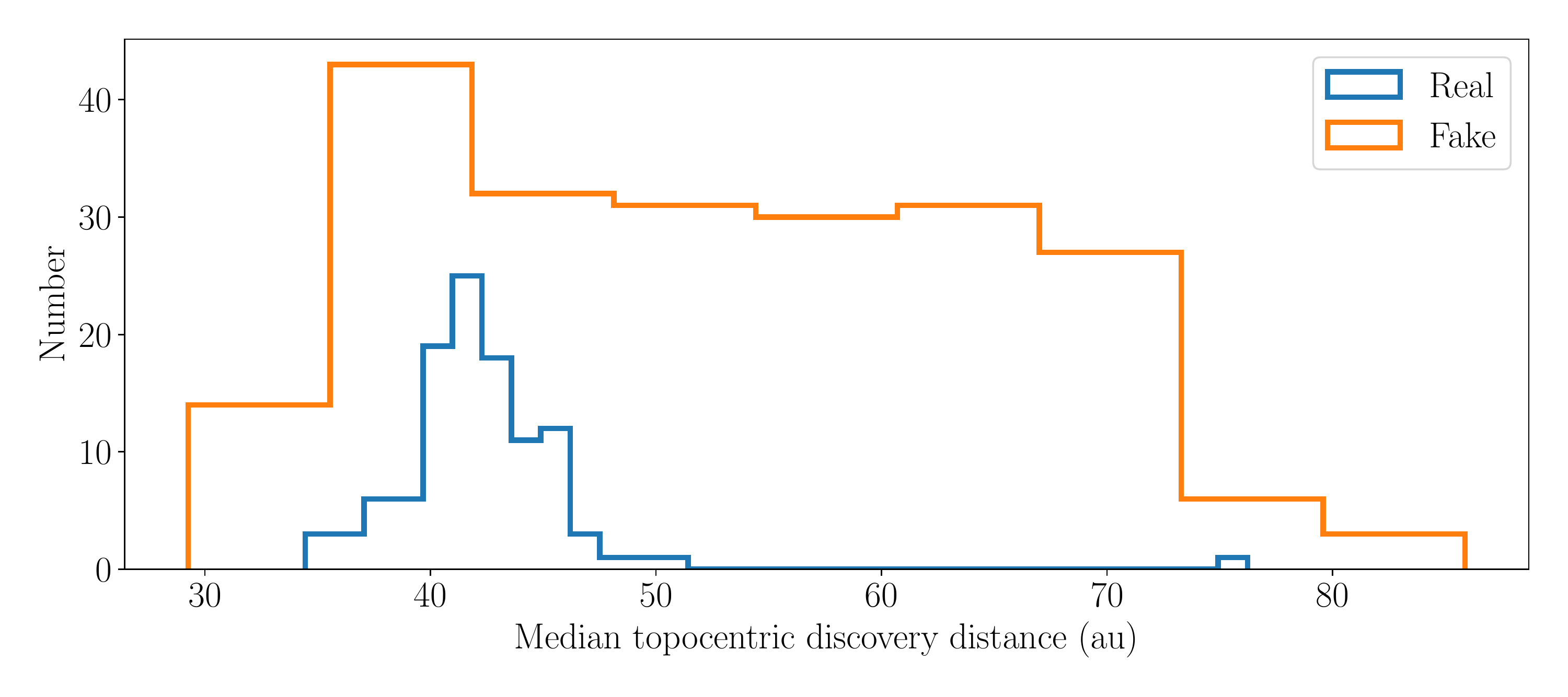}
    \caption{The median topocentric discovery distance for all correctly-linked fake orbits and all 110 real orbits that pass all orbital cuts. The median is taken from the distribution of topocentric distances for each long stare (and thus each tracklet) in the linked orbit.}
    \label{fig:discovery_distance}
\end{figure}

In Figure \ref{fig:discovery_distance}, we show the median topocentric discovery distance for the 230 correctly-linked fake orbits and the 110 real orbits which pass all orbital cuts. This distance is defined as the median of the topocentric distances reported by the orbit fit at each long stare for which the object was observed. As expected, our results for the fake objects are consistent with the distances for the fake objects detected in single long stares, shown in Figure 1 of \cite{DEEPIII}. The distribution of discovery distances for the fake orbits encompasses the distribution of the discovery distances of the 110 real orbits, as one would expect for correctly-linked real orbits.

In order to determine dynamical classifications for our objects, we follow the classification scheme of \citet{Gladman_2008} and \cite{Khain_2020} using the implementation of \cite{Bernardinelli2022}. We create 20 clones of each object, and then use \texttt{Rebound} \citep{Rein_2012} with the \texttt{WHFast} symplectic integrator \citep{Wisdom1991,Rein_2015} to integrate these for 10 Myr from the starting time $t_0$. Clones that librate around the resonance argument for more than 90\% of the simulation time are considered resonant. Clones where the semi-major axis varies by more than 3.75\% from $a(t_0)$ are considered scattering. Clones where $a(t_0)<a_N$ are considered inner Centaurs, where $a_N$ is the semi-major axis of Neptune. Clones with $e(t_0)>0.24$ are considered detached. Remaining clones are considered classicals.

\begin{figure}[tbh]
    \centering
    \includegraphics[width=\textwidth]{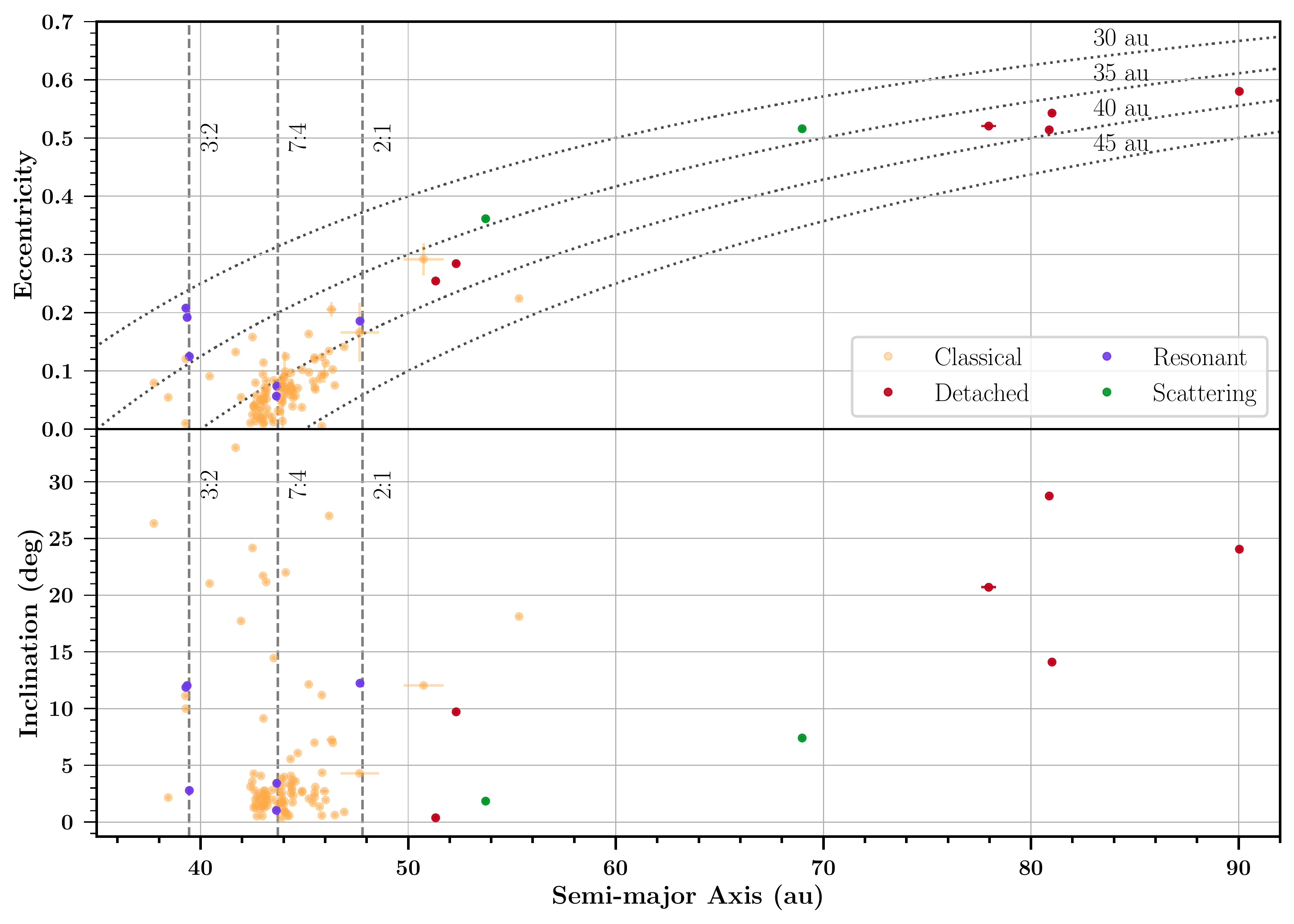}
    \caption{Eccentricity versus semi-major axis (top) and inclination versus semi-major axis (bottom) for the 110 objects from the DEEP B1 fields. Color denotes dynamical classifications. Vertical dotted lines correspond to the approximate location of select Neptunian $p$:$q$ mean motion resonance. Curved dotted lines (top panel only) correspond to lines of constant pericenter.}
    \label{fig:Dynamical_Classification}
\end{figure}

Objects are classified based on the classification of the majority of their clones ($>50\%$). Objects where more than 80\% of the clones show the same mean-motion resonance (MMR) for more than 90\% of the time are classified as securely resonant, and those with $50-80$\% or more of the clones showing resonant behavior are classified as insecurely resonant. Objects where no single classification reaches a majority are considered insecurely classified; these are presented in Table \ref{tb:objects}. All 110 TNOs and their classifications are shown in Figure \ref{fig:Dynamical_Classification}. 

We identify six resonant TNOs, in the 3:2, 7:4, 2:1 resonances, with more objects in the 3:2. This is in line with previous surveys (e.g. \citealt{Bernardinelli2022}), but we will defer any statistical claims to future releases, when fields that are expected to contain more resonant objects are analyzed. The B1 fields are close to Neptune's position in the sky, so these fields are generally far from the 3:2, 7:4, and 2:1 resonant populations, which explains the proportionately-smaller number of resonant objects compared to (for example) \cite{Bannister_2018} or \cite{Bernardinelli2022}.

We identify 96 classical TNOs, the majority of our sample, unsurprisingly, given our survey geometry. 66 of these objects are dynamically ``cold'' ($i \leq 5\deg$), another feature expected from the survey geometry, as our fields are in the invariable plane. The four recovered known objects are in this population. One of our ``classical'' objects has $a \approx 56$ au, close to the 5:2 MMR, adding to the population of high-$q$ non-resonant objects near these MMRs \citep[see][]{Sheppard2016,lawler2019,Bernardinelli2022}.

We also identify six detached objects. One of these detached objects ($a=77.53$ au, $e=0.52$, $i=20.69^\circ$) was detected with a median topocentric discovery distance of 76.23 au, being our most distant object discovery. Finally, we also have two scattering objects. With the detection of these scattering objects, we demonstrate that we are able to detect and characterize the four main dynamical populations of TNOs (classical, scattering, detached, and resonant).

Finally, we investigate the observed $H$ magnitude distributions. We compare the residuals of the simulated values for median $H$ magnitude and the median of the observed values for $H$ magnitude. Note that, as the fakes were generated without phase coefficients, we likewise do not include these here. The median of the residuals in $H$ is $-0.001$, indicating that there is no observable systematic offset. The standard deviation is $\sigma=0.086$ ($\sigma_G=0.031$), also indicating that we are correctly recovering our absolute magnitudes.

\begin{figure}[h]
    \centering
    \includegraphics[width=0.7\textwidth]{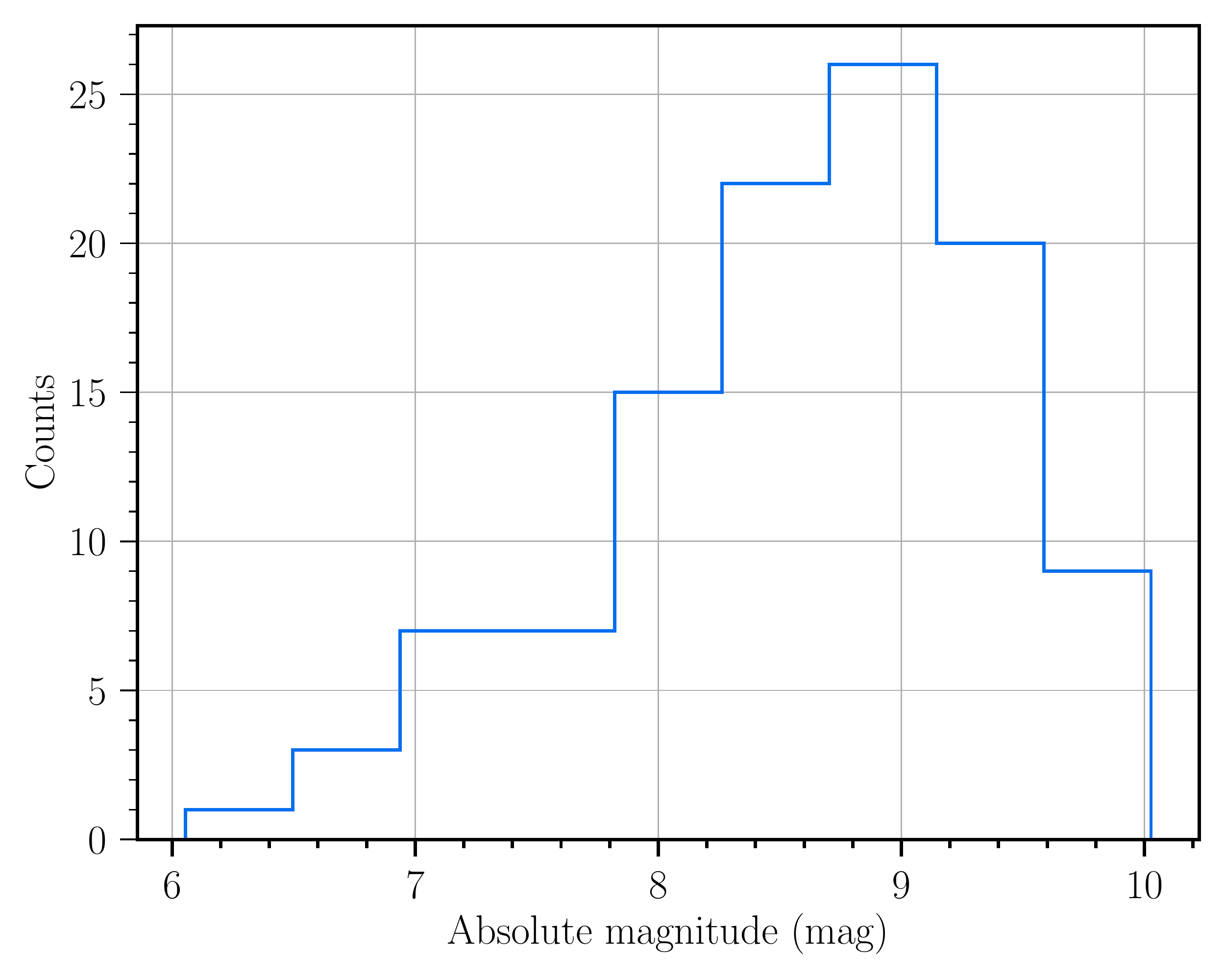}
    \caption{The inverse variance weighted mean H magnitudes of the 110 detected real objects. This is computed using an H-G model with an assumed value of $G=0.15$.}
    \label{fig:DEEP_H_Mags}
\end{figure}

As our real objects have between 4 and 6 shift and stack detections with similarly low phase angles (as expected from the survey design), and are in their majority cold Classicals, a population that has higher variability than the other TNO subpopulations \citep{DEEPIV,bernardinelli2023}, we elect to report the mean absolute magnitude $\overline{H}$ over all detections; any choice of phase coefficient and typical light curve amplitude would lead to systematic offsets in $H$ for an undetermined subset of the population (for which these parameters are not representative). We show the distribution of these values in Figure \ref{fig:DEEP_H_Mags}. Our smallest object (that is, the one with the largest $H$) has $\overline{H} = 10.2$, being one of the smallest objects with $q>30$ au with multi-year arcs to date. 

\subsection{Structure of the classical Kuiper belt}\label{sec:kernel}
We investigate the distribution of semi-major axis of our classical TNOs and their relationship to models of the Kuiper belt. We compare our data to the L7 model of the Canada–France Ecliptic Plane Survey \citep[CFEPS,][]{Petit_2011}, where the classical Kuiper belt is divided into a dynamically excited, ``hot'' population, and two dynamically cold ones: the ``kernel'', concentrated in the $43.8 < a < 44.8$ au range, and the ``stirred'' population covering the $40 < a < 48$ au range. While this model is known to not reproduce the inclination distribution of these objects \citep{Petit2017, Bernardinelli2022}, this model reproduces well their semi-major axis distribution \citep{Bannister2015,Bernardinelli2022}. Since we are probing a different size range than the one in the published CFEPS-L7 model, we have re-sampled the model and changed the size distribution of the cold and hot classical populations to those of \cite{Fraser2014}, to provide a sufficient sample of objects in our size range. 

\begin{figure}[h]
    \centering
    \includegraphics[width=0.7\textwidth]{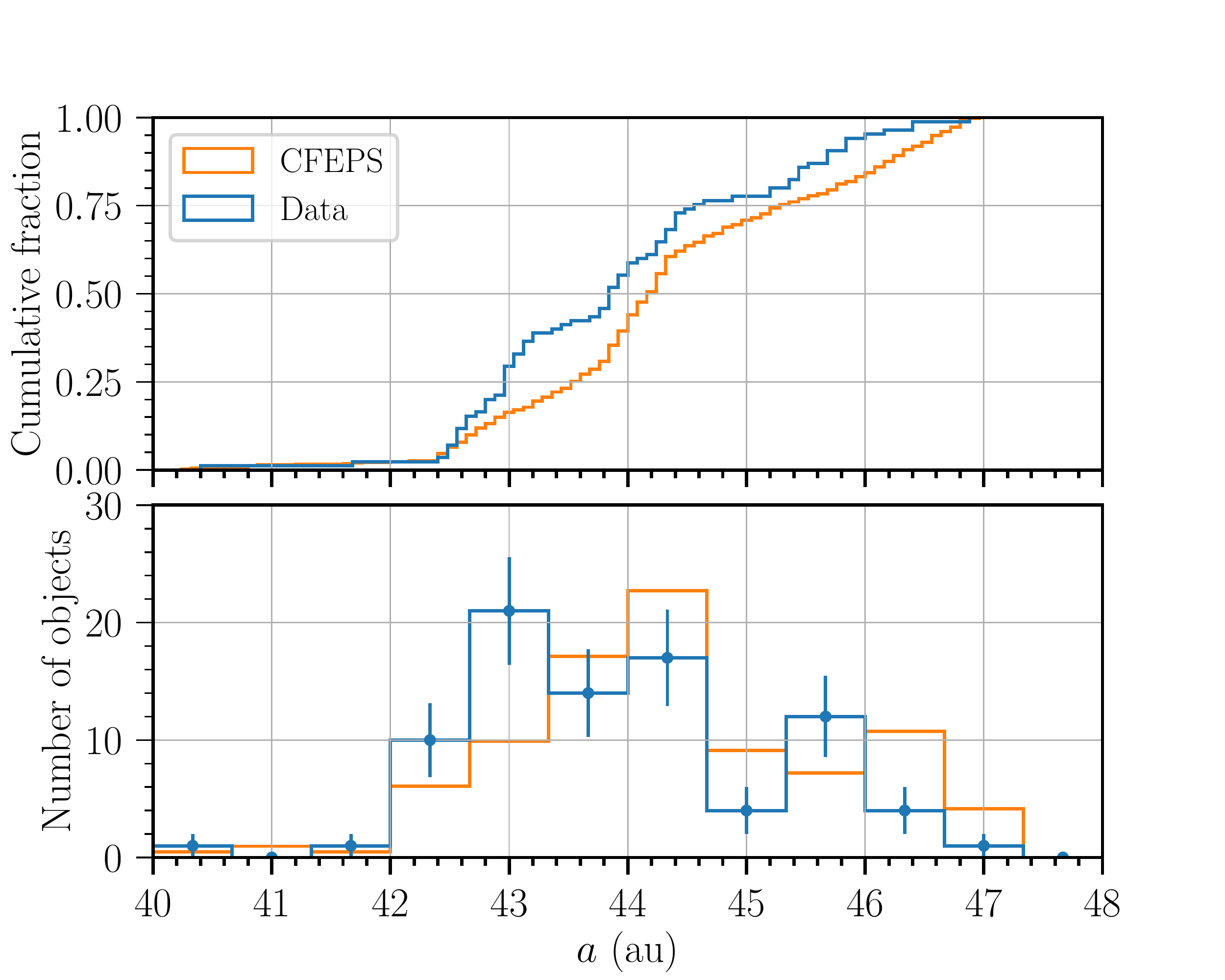}
    \caption{Cumulative (top) and differential (bottom) histograms of semi-major axis distribution of the predicted CFEPS detections (orange) given our survey biases and our recovered classical objects (blue), limited to the $40 < a < 48$ au range. The differential histogram was normalized to the number of real objects, and we also include Poisson error bars in the counts of the real objects.}
    \label{fig:kernel}
\end{figure}

We have used our survey simulator \citep{DEEPIII} to produce a population of objects based on this model when subject to our survey biases, and compare their $a$ distribution with our data. The results are shown in Figure \ref{fig:kernel}. We do not see a sharp increase in the number of objects at $43.8 < a <  44.8$ au predicted by the narrow Kuiper belt kernel in the CFEPS-L7 model. The model distribution increases sharply in the $42.6 \lesssim a \lesssim 44.6$ range, and systematically over-predicts objects with $a \gtrsim 46$ au when compared to the observed data. We compute a Kolmogorov-Smirnov (KS) test \citep{Press2007} between these two distributions, and obtain a $p$-value $\leq 10^{-5}$ that the data and the model are drawn from the same distribution, confirming the visual impression that these distributions do not agree.

An implication here is that the kernel either must be wider than the $1$ au currently in the CFEPS-L7 model, since our object density is $\sim$constant (within $1\sigma$) in the $42.6$ au to $44.6$ au range, or that the cold Classical population must have a different structure than that implied by the stirred+kernel model. A more complicated alternative would be for such differences to be as a function of size: the CFEPS-L7 model was derived with $H < 8$ objects, while here the majority of our objects are fainter than such $H$. However, by restricting ourselves to $H < 8$ in both the model and data, the K-S test leads to a $p$-value $=0.0047$, which still indicates a discrepancy. Understanding these differences will have implications for dynamical models that seek to reproduce the Kuiper belt. In particular, the formation of the kernel requires, for example, that Neptune ``jumps'' during its migration \citep{Nesvorny2015}. We defer more detailed tests of these dynamical models to future releases of the DEEP data, when our object counts are expected to increase by a factor of 10 \citep{DEEPI}.

We also note that \cite{DEEPV} used the DEEP B1 data to infer the absolute magnitude distribution for the cold Classical population. Further analysis of DEEP data will, then, enable more detailed tests of the joint radial and absolute magnitude distribution of the Kuiper belt.

\section{Summary} \label{sec:discuss}

We have demonstrated that we are able to systematically link and fit high-quality multi-year orbits to sources identified with digital tracking. We have recovered 110 TNOs (4 already known), including objects as faint as $m_{VR} \approx 26.2$ and $H_{VR} \approx 10.2$. This means that we are able to probe the dynamical structure of the trans-Neptunian region to systematically smaller sizes than any other TNO survey with more than 100 objects to date. The population is, as expected from the survey design and geometry, dominated by classical Kuiper belt objects in the $40 < a < 48$ au range, but we have also recovered a detached object ($a \approx 78$ au) with $H_{VR} \approx 6.1$, and a few TNOs in the other dynamical classes. We used our discovered Kuiper objects and our survey simulator to compare the distribution of our classical objects with the CFEPS-L7 model, and conclude that our data is inconsistent with that model. Future investigations of the Kuiper belt should lead to improved model constraints of this region. 

 As this is the first multi-year arc release of the DEEP survey, a number of improvements are expected, including a more complete astrometric characterization and an additional year of observations for the B1 field. We also have more fields in which this analysis will be conducted. We can expect that the number of discovered TNOs by the DEEP survey will significantly increase, and this increase in statistical power will make these data set even more valuable for testing models of the trans-Neptunian region.

\software{KBMOD \citep{kbmod}, LSST Science Pipelines \citep{LSST_DM}, astropy \citep{astropy}, scikit-image \citep{skimage}, numpy \citep{numpy}, CUDA \citep{cuda}, scikit-learn \citep{scikit}, pandas \citep{pandas}, matplotlib \citep{matplotlib}, tensorflow \citep{tensorflow}, orbitspp \citep{Bernstein2000}, deslinker \citep{Bernardinelli2019,Bernardinelli2022}, vaex \citep{Breddels2018}}

\bibliographystyle{aasjournal}
\bibliography{references}

\begin{acknowledgements}
This work is based in part on observations at Cerro Tololo Inter-American Observatory at NSF’s NOIRLab (NOIRLab Prop. ID 2019A-0337; PI: D. Trilling), which is managed by the Association of Universities for Research in Astronomy (AURA) under a cooperative agreement with the National Science Foundation.

This work is supported by the National Aeronautics and Space Administration under grant No.\ NNX17AF21G issued through the SSO Planetary Astronomy Program and by the National Science Foundation under grants No.\ AST-2009096 and AST-2107800. This research was supported in part through computational resources and services provided by Advanced Research Computing at the University of Michigan, Ann Arbor. This work used the Extreme Science and Engineering Discovery Environment \citep[XSEDE; ][]{XSEDE}, which is supported by National Science Foundation grant number ACI-1548562. This work used the XSEDE Bridges GPU and Bridges-2 GPU-AI at the  Pittsburgh Supercomputing Center through allocation TG-AST200009.

H. Smotherman acknowledges support by NASA under grant No.\ 80NSSC21K1528 (FINESST). H. Smotherman, M. Juri\'{c} and P. Bernardinelli acknowledge the support from the University of Washington College of Arts and Sciences, Department of Astronomy, and the DiRAC Institute. The DiRAC Institute is supported through generous gifts from the Charles and Lisa Simonyi Fund for Arts and Sciences and the Washington Research Foundation. M. Juri\'{c} wishes to acknowledge the support of the Washington Research Foundation Data Science Term Chair fund, and the University of Washington Provost’s Initiative in Data-Intensive Discovery. 

This project used data obtained with the Dark Energy Camera (DECam), which was constructed by the Dark Energy Survey (DES) collaboration. Funding for the DES Projects has been provided by the US Department of Energy, the US National Science Foundation, the Ministry of Science and Education of Spain, the Science and Technology Facilities Council of the United Kingdom, the Higher Education Funding Council for England, the National Center for Supercomputing Applications at the University of Illinois at Urbana-Champaign, the Kavli Institute for Cosmological Physics at the University of Chicago, Center for Cosmology and Astro-Particle Physics at the Ohio State University, the Mitchell Institute for Fundamental Physics and Astronomy at Texas A\&M University, Financiadora de Estudos e Projetos, Funda\c{c}\~{a}o Carlos Chagas Filho de Amparo \'{a} Pesquisa do Estado do Rio de Janeiro, Conselho Nacional de Desenvolvimento Cient\'{i}fico e Tecnol\'{o}gico and the Minist\'{e}rio da Ci\^{e}ncia, Tecnologia e Inova\c{c}\~{a}o, the Deutsche Forschungsgemeinschaft and the Collaborating Institutions in the Dark Energy Survey.

The Collaborating Institutions are Argonne National Laboratory, the University of California at Santa Cruz, the University of Cambridge, Centro de Investigaciones En\'{e}rgeticas, Medioambientales y Tecnol\'{o}gicas–Madrid, the University of Chicago, University College London, the DES-Brazil Consortium, the University of Edinburgh, the Eidgen\"{o}ssische Technische Hochschule (ETH) Z\"{u}rich, Fermi National Accelerator Laboratory, the University of Illinois at Urbana-Champaign, the Institut de Ci\`{e}ncies de l’Espai (IEEC/CSIC), the Institut de Física d’Altes Energies, Lawrence Berkeley National Laboratory, the Ludwig-Maximilians Universit\"{a}t M\"{u}nchen and the associated Excellence Cluster Universe, the University of Michigan, NSF’s NOIRLab, the University of Nottingham, the Ohio State University, the OzDES Membership Consortium, the University of Pennsylvania, the University of Portsmouth, SLAC National Accelerator Laboratory, Stanford University, the University of Sussex, and Texas A\&M University.

\end{acknowledgements}

\end{document}